\documentclass[useAMS,usenatbib]{mn2e}
\usepackage{epsfig}
\usepackage{amsmath, amssymb,bm}

\title[On Orbital Period Changes in Nova Outbursts
]{On Orbital Period Changes in Nova Outbursts }

\author[R. G. Martin, M. Livio \& B. E. Schaefer]{Rebecca
  G. Martin\thanks{E-mail: rmartin@stsci.edu}$^1$, Mario Livio$^1$ and
  Bradley E. Schaefer$^2$ \\ $^1$Space Telescope Science Institute,
  3700 San Martin Drive, Baltimore, MD 21218 \\ $^2$Physics and
  Astronomy, Louisiana State University, Baton Rouge, LA 70803 \\}

\begin{document}

\date{}

\pagerange{\pageref{firstpage}--\pageref{lastpage}} 
\pubyear{2010}
\maketitle

\label{firstpage}

\begin{abstract}

We propose a new mechanism that produces an orbital period change
during a nova outburst. When the ejected material carries away the
specific angular momentum of the white dwarf, the orbital period
increases. A magnetic field on the surface of the secondary star
forces a fraction of the ejected material to corotate with the star,
and hence the binary system. The ejected material thus takes angular
momentum from the binary orbit and the orbital period decreases.  We
show that for sufficiently strong magnetic fields on the surface of
the secondary star, the total change to the orbital period could even
be negative during a nova outburst, contrary to previous expectations.
Accurate determinations of pre- and post-outburst orbital periods of
recurrent nova systems could test the new mechanism, in addition to
providing meaningful constraints on otherwise difficult to measure
physical quantities. We apply our mechanism to outbursts of the
recurrent nova U Sco.

\end{abstract}


\section{Introduction}

Classical novae are binary systems in which mass is transferred from a
main-sequence star on to a white dwarf by Roche-lobe overflow. The
critical amount of mass that can be accreted on to the surface of the
white dwarf prior to an outburst is a strongly decreasing function of
the white dwarf mass \citep{truran86}.  At this mass limit, the
temperature and density at the base of the accreted layer are high
enough for hydrogen to ignite. The temperature then rises rapidly in a
thermonuclear runaway \citep{starrfield88} and the pressure at the
base of the accreted layer becomes high enough that the accreted mass
(and sometimes a little more) is ejected \citep[e.g.][]{kovetz85}.

Recurrent novae show outbursts at intervals of $10-80\,\rm yr$
\citep{warner95, webbink87}. To account for the short timescale
between the outbursts, the white dwarf in a recurrent nova system must
have a mass, $M_1$, close to the Chandrasekhar limit
\citep[e.g.][]{kato88,kato89}. We consider in more detail the
recurrent nova U Sco that most recently erupted in 2010. The evolved
companion in U Sco has a mass of $M_2=0.88\,\rm M_\odot$ so the mass
ratio, $q=M_2/M_1=0.64$, is relatively large.

In Section~\ref{old} we begin by re-examining all the previously
proposed sources of orbital period change during a nova outburst. We
first describe a simple model where the material carries away its
specific angular momentum, then we include mass accretion on to the
companion and frictional angular momentum losses as the binary moves
through the common envelope. In Section~\ref{new} we propose a new
mechanism for orbital period change involving the magnetic field on
the secondary star. In Section~\ref{usco} we apply our model to the
outbursts in the recurrent nova U Sco.

\section{Outburst Model}
\label{old}

We consider first a simple model of the outburst where the ejected
material carries away the specific angular momentum of the white
dwarf.  The non-degenerate mass accumulated on to the surface of the
white dwarf is very thin.  The envelope is ejected when the pressure
at the surface of the white dwarf reaches a critical value of the
order of $P_{\rm crit}=10^{20}\,\rm dyn\,cm^{-2}$
\citep[e.g.][]{fujimoto82a,fujimoto82b,macdonald83}. The critical
amount of mass that accumulates before a nova outburst is of order
\begin{equation}
\Delta m_1=4\pi R_1^4 \frac{P_{\rm crit}}{G M_1},
\label{dm}
\end{equation}
for a white dwarf of mass $M_1$ and radius $R_1$.  Both theory and
observations suggest that all the the material that has been accreted
since the last outburst is ejected in the outburst

The angular momentum of the binary star system is given by
\begin{equation}
J=\frac{M_{2} M_1}{M}a^2 \Omega,
\label{da}
\end{equation}
where $a$ is the separation of the two stars and the angular velocity,
$\Omega$, is given by Kepler's law
\begin{equation}
\Omega^2=\frac{GM}{a^3}.
\label{kepler}
\end{equation}
Here the mass of the companion star is $M_2$ and the total mass of the
system is $M=M_1+M_2$.  

If the ejected mass carries away its specific angular momentum and all
the mass is lost from the system, then the angular momentum removed from
the system is
\begin{equation}
\Delta J_{\rm spec}=-\Delta m_1 \,a_1^2 \Omega,
\label{dj}
\end{equation}
where $a_1$ is the distance of $M_1$ to the centre of mass of the binary
\begin{equation}
a_1=\frac{M_{2}}{M}a.
\label{a1}
\end{equation}
We neglect the spin of the white dwarf because it is not coupled to
the binary orbit.  By differentiating equations~(\ref{da})
and~(\ref{kepler}) we find the change in angular momentum of the
binary
\begin{equation}
\frac{\Delta J}{J}=\frac{\Delta M_1}{M_1}+\frac{\Delta M_2}{M_2}-\frac{1}{2}\frac{\Delta M}{M}+\frac{1}{2}\frac{\Delta a}{a},
\label{change}
\end{equation}
where $\Delta a$ is the corresponding change in the separation of the
system (due to mass lost) during the outburst, $\Delta M_1$ is the
change in mass of the white dwarf, $\Delta M_2$ is the change in the
mass of the secondary and $\Delta M$ is the change to the total mass
of the system.  For the case where all the ejected material is lost
from the system, we choose $\Delta M_1=-\Delta m_1$ and
$\Delta M_2=0$.  We take the angular momentum loss from the system to
be equal to that carried away by the ejected mass so that $\Delta
J=\Delta J_{\rm spec}$.  With equations~(\ref{da}),~(\ref{dj})
and~(\ref{change}) we find
\begin{equation}
\frac{\Delta a}{a}=\frac{\Delta m_1}{M}.
\label{over}
\end{equation}
As mass is lost from the system in the outburst the separation
increases. 

\begin{figure}
\begin{center}
\includegraphics[width=8.4cm]{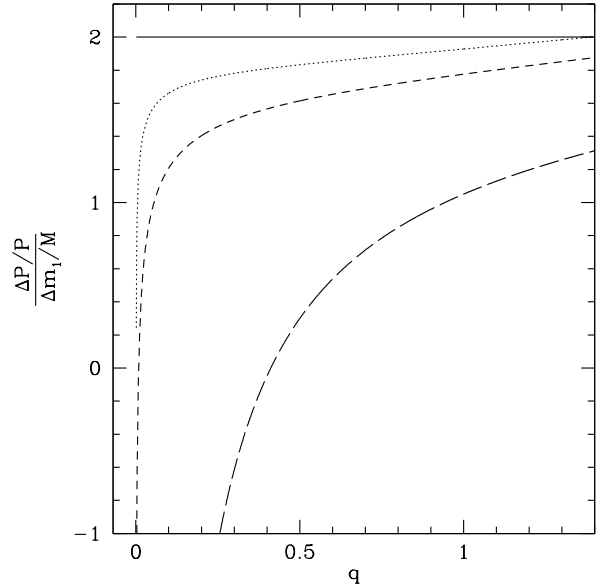}
\end{center}
\caption[]{The change $\frac{\Delta P/P}{\Delta m_1/M}$ as a function
  of mass ratio, $q=M_2/M_1$. For the solid line the ejected mass just
  carries away its specific angular momentum
  (equation~\ref{eqdp}). The dotted line shows the case where
  accretion onto the secondary is considered (corresponding separation
  change in equation~\ref{mass}). The short dashed line includes
  accretion on to the secondary and frictional angular momentum losses
  (corresponding separation change in equation~\ref{fric}). The long
  dashed line takes into account the angular momentum change of the
  ejected material due to a magnetic field on the secondary that has
  $R_{\rm A}=0.75\,a$ (corresponding separation change in
  equation~\ref{cor}). We calculate the period change from the
  separation change with equation~(\ref{dp}).}
\label{dadm}
\end{figure}

By differentiating equation~(\ref{kepler}) we find the period change
during the outburst to be
\begin{equation}
\frac{\Delta P}{P}=-\frac{\Delta \Omega}{\Omega}
=-\frac{1}{2}\frac{\Delta M}{M}+\frac{3}{2}\frac{\Delta a}{a}
\label{dp}
\end{equation}
and, for the case where the material carries away all of its specific
angular momentum,with equation~(\ref{over}) we find
\begin{equation}
\frac{\Delta P}{P}=2\frac{\Delta m_1}{M}.
\label{eqdp}
\end{equation}
Since $\Delta m_1>0$ we see that the period of the system should
increase during the outburst if the material carries away its specific
angular momentum. In Fig.~\ref{dadm} we plot $\frac{\Delta P/P}{\Delta
  m_1/M}$ as a function of the mass ratio, $q$, for the different
outburst models we consider. The solid constant line corresponds to
this model where the mass ejected carries away its specific angular
momentum.

\subsection{Mass Accretion on to the Companion}

Now we consider the effect of a fraction of the ejected mass, $\beta$,
that may be captured by the companion in the outburst. We take $\Delta
M_1=-\Delta m_1$ and $\Delta M_2=\beta \Delta m_1$ so that 
\begin{equation}
\Delta J=-(1-\beta)\Delta m_1\,a_1^2\Omega.
\end{equation}
 The corresponding change to the separation is then
\begin{equation}
\frac{\Delta a}{a}=\frac{\Delta m_1}{M} \left[1+\beta
  \left(2q-1-\frac{2}{q}\right) \right]
\label{mass}
\end{equation}
\citep{shara86}, where $q=M_{2}/M_1$. This reduces to
equation~(\ref{over}) with $\beta=0$. In the absence of strong
magnetic effects the maximum value of $\beta$ is the fractional area
of the companion's radius. We can estimate the captured fraction of
mass with
\begin{equation}
\beta=\frac{\pi R_2^2}{4 \pi a^2}.
\end{equation}
Because the secondary fills its Roche lobe, we can estimate the
stellar radius with $R_2=R_{\rm L}$, where
\begin{equation}
R_{\rm L}=a\left(\frac{0.49\,q^{\frac{2}{3}}}{0.6\,q^\frac{2}{3}+\ln(1+q^\frac{1}{3})}\right)
\label{rl}
\end{equation}
\citep{eggleton83}.  We compute the period change for this model
including mass accretion on to the secondary star with
equations~(\ref{dp}) and~(\ref{mass}). In Fig.~\ref{dadm} we plot
$\frac{\Delta P/P}{\Delta m_1/M}$ as a function of the mass ratio for
this model (dotted line). Generally, the separation change is not so
large as the previous case where all the ejected mass left the binary
system. However, for the largest mass ratios the period increases
compared to the previous model.

\subsection{Frictional Angular Momentum Losses}

Angular momentum is lost from the binary system because of frictional
angular momentum losses as the binary moves through the common
envelope created by the ejected material
\citep[e.g.][]{macdonald80,macdonald86,shara86,livio90}. This causes
the separation of the system to decrease and so the period decreases
too. They found that the separation change given in
equation~(\ref{mass}) becomes
\begin{align}
\frac{\Delta a}{a}= & \frac{\Delta m_1}{M} \left\{1+\beta
  \left[2q-1-\frac{2(1+ x)}{q}\right]\right\}
\label{fric}
\end{align}
\citep{shara86}, where $x=\sqrt{2}$ for a very slow nova and $x=1$ for
a fast nova.  We note that \cite{livio91} suggest that assumptions
made in deriving this mean that is is valid only for novae with longer
orbital periods. Again, we can find the period change for this model
with equation~(\ref{dp}). In Fig.~\ref{dadm} we plot $\frac{\Delta
  P/P}{\Delta m_1/M}$ as a function of the mass ratio for this model
for a slow nova (short-dashed line). The separation change (and hence
period change) here is, as expected, less than that for the
previous case without the frictional angular momentum losses.

\subsection{Magnetic Braking and Gravitational Radiation}

There are also two mechanisms that remove angular momentum from the
system between outbursts that we consider briefly here.  For
systems with relatively long orbital periods ($P \gtrsim 3\,\rm hr$),
magnetic braking provides the largest continual loss of angular
momentum from the system. The rate of loss of angular momentum is
given roughly by
\begin{equation}
\dot J_{\rm MB}=-5.83\times 10^{-16}\left(\frac{R_1}{R_\odot}\right)^3 (\Omega \,\rm yr)^3\,\rm M_\odot R_\odot^2\,yr^{-2}
\end{equation}
\citep{rappaport83}.  For reasonable parameters this gives a timescale
of around a few billion years \citep[e.g.][]{martin05}.

For the closest systems ($P \lesssim 3\,\rm hr$), gravitational
radiation is the dominate cause of angular momentum loss from the
binary system between outbursts. The rate of loss of angular momentum
from two point masses in a circular orbit is
\begin{equation}
\dot J_{\rm GR}=-\frac{32G^3}{5c^5}\frac{M_1M_2M}{a^4}
\end{equation}
\citep{landau51}. This gives a timescale of order several million
years except for the very closest of systems. Neither gravitational
radiation nor magnetic braking will have any effect on a system on an
observable timescale.

\section{Corotating Ejected Material}
\label{new}

\begin{figure}
\begin{center}
\includegraphics[width=8.4cm]{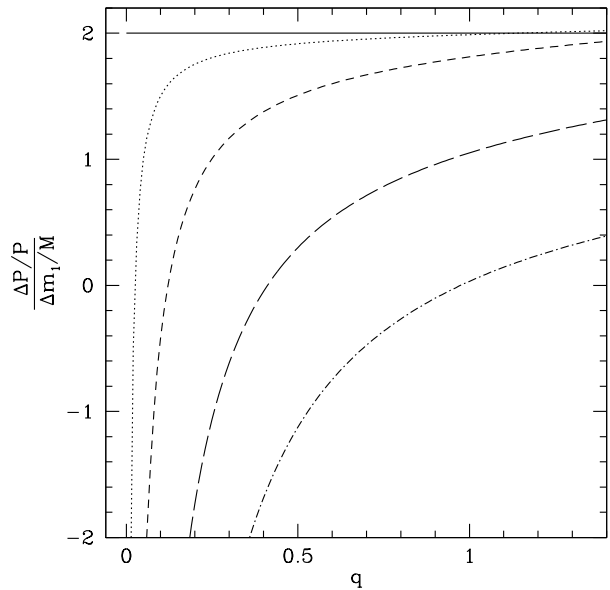}
\end{center}
\caption[]{The change $\frac{\Delta P/P}{\Delta m_1/M}$ as a function
  of mass ratio, $q$. For the solid line the ejected mass carries away
  only its specific angular momentum. All other lines take into
  account the angular momentum change of the ejected material due to a
  magnetic field on the secondary (corresponding separation change in
  equation~\ref{cor}). In order of decreasing height the lines have
  $R_{\rm A}=0.25\,a$ (dotted line), $R_{\rm A}=0.5\,a$ (short-dashed
  line), $R_{\rm A}=0.75\,a$ (long dashed line) and $R_{\rm A}=0.9\,a$
  (dot-dashed line).}
\label{dadm2}
\end{figure}

In this Section we consider an entirely new mechanism that can change
the orbital period during a nova outburst. Suppose ejected material
that moves within a critical radius (that depends on the magnetic
field strength) of the secondary star couples to its magnetic field
and so is forced to corotate with the binary orbit. The transfer of
angular momentum between the ejected material and the binary orbit
causes a change to the orbital period of the system.

The magnitude of the dipole magnetic field of the secondary is
\begin{equation}
B=\frac{\mu}{R^3},
\end{equation}
where $\mu$ is the dipole moment and $R$ is the distance to the
secondary star.  The magnetic field energy density is
\begin{equation}
E_{\rm mag}=\frac{B^2}{4\pi}.
\end{equation}
The kinetic energy density of the ejected matter is
\begin{equation}
E_{\rm kin}= \frac{1}{2}\rho u^2,
\end{equation}
where $\rho$ is the density of the ejected material. The ejection
velocity, $u$, is in the range $300 - 3000\,\rm km\,s^{-1}$
\citep{shara86}.  From the continuity equation, close the the
secondary star, we approximately have
\begin{equation}
\rho u=\frac{\dot M}{4 \pi a^2},
\end{equation}
where $\dot M$ is the mass ejection rate from the primary white dwarf
star. We approximate the average
mass-loss rate with
\begin{equation}
\dot M=\frac{\Delta m_1}{\tau},
\label{mdot}
\end{equation}
where $\tau$ is the timescale over which the mass is lost.

The magnetic energy density is equal to the kinetic energy density at
the Alfv\'{e}n radius which we find to be
\begin{equation}
R_{\rm A}=\left(\frac{2 \mu^2 a^2}{\dot M u}\right)^\frac{1}{6}.
\label{ra}
\end{equation}
We consider this further in Section~\ref{usco} where we apply our
model to the recurrent nova U Sco.  We parametrise the dipole moment
with the magnetic field strength at the stellar surface, $B_{\rm
  star}$, so that
\begin{equation}
\mu=B_{\rm star}R_2^3,
\end{equation}
where $R_2$ is the stellar radius of the companion which we find with
equation~(\ref{rl}).  

The specific angular momentum of the ejected mass that is forced to
corotate with the secondary is $(a_2^2+KR_{\rm A}^2)\,\Omega$ assuming
that the secondary is tidally locked.  The distance from the centre of
mass to the secondary star is
\begin{equation}
a_2=\frac{M_1}{M}a.
\end{equation}
The constant $K$ depends on the distribution of the material within
the Alfv\'{e}n radius. If it were distributed uniformly within a
spherical shell, then $K=2/3$. Because the density within the shell
varies and the shell itself will not be perfectly spherical, we take
$K=1$. Thus, the angular momentum loss from the binary to the ejected
material is given by
\begin{equation}
\Delta J_{\rm cor}= -f \Delta m_1 (a_2^2+R_{\rm A}^2-a_1^2)\Omega.
\label{jobs}
\end{equation}
We estimate the fraction of the ejected mass that gains angular
momentum in this way to be
\begin{equation}
f=\frac{R_{\rm A}^2}{4 a^2}.
\label{eqf}
\end{equation}
Now with equations~(\ref{da}) and~(\ref{jobs}) we find
\begin{equation}
\frac{\Delta J_{\rm cor}}{J}=-\frac{f\Delta m_1}{M}\frac{ (1+q)^2}{q}\left[
\left(\frac{R_{\rm A}}{a}\right)^2+\frac{1-q}{1+q}
\right]  .
\end{equation}
We substitute this into equation~(\ref{change}) with 
\begin{equation}
\Delta J=\Delta J_{\rm spec}+\Delta J_{\rm cor},
\end{equation}
$\Delta M_1=-\Delta m_1$ and $\Delta M_2=0$ to find the change in
separation
\begin{equation}
\frac{\Delta a}{a}=\frac{\Delta m_1}{M}\left\{ 1-2f
\frac{ (1+q)^2}{q}\left[
\left(\frac{R_{\rm A}}{a}\right)^2+\frac{1-q}{1+q}
\right] 
\right\}.
\label{cor}
\end{equation}
We note that this reduces to equation~(\ref{over}) when there is no
magnetic field.  In Fig.~\ref{dadm2} we plot $\frac{\Delta P/P}{\Delta
  m_1/M}$ as a function of the mass ratio for different values of the
Alfv\'{e}n radius.  The larger magnetic fields are even capable of
producing an overall decrease to the orbital period during the nova
for small mass ratios.

In Fig.~\ref{dadm} we also plot $\frac{\Delta P/P}{\Delta m_1/M}$ for
an Alfv\'{e}n radius of $0.75\,a$ (long dashed line) for comparison
with the other models we have considered. For this strong magnetic
field, the period change is smaller than with mass transfer to the
secondary or frictional angular momentum losses. If a secondary star
has a large magnetic field then it will significantly alter the
orbital period change during a nova outburst.

\begin{figure}
\begin{center}
\includegraphics[width=8.4cm]{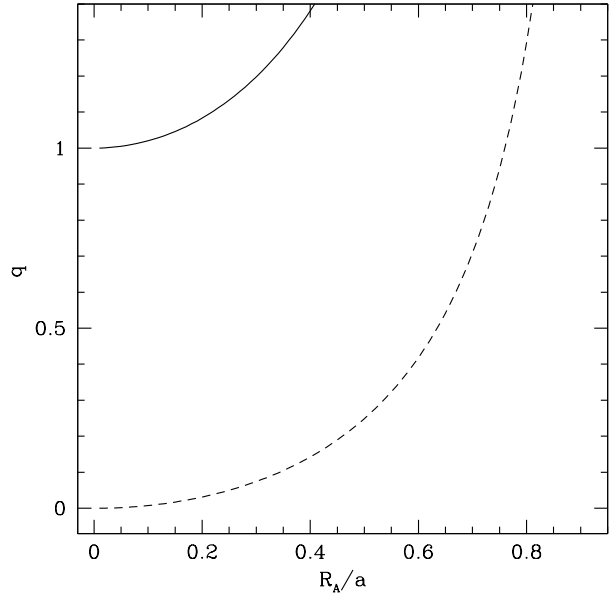}
\end{center}
\caption[]{The solid line shows the critical mass ratio, $q_{\rm
    crit}$, as a function of the Alfv\'{e}n radius. If $q<q_{\rm
    crit}$ the period decreases due to the magnetic field, whereas if
  $q>q_{\rm crit}$ the period increases. Below the dashed line, the
  period change becomes negative.}
\label{qc}
\end{figure}

There is a critical mass ratio where the additional term in
equation~(\ref{cor}) causes no change to the separation (or orbital
period)
\begin{equation}
q_{\rm crit}= \frac{1+\left(\frac{R_{\rm a}}{a}\right)^2}{1-\left(\frac{R_{\rm a}}{a}\right)^2}  .
\end{equation}
If $q<q_{\rm crit}$ then the ejected material that is forced to
corotate with the secondary gains angular momentum, angular momentum
is lost from the orbit and so the separation decreases.  Similarly the
orbital period change is smaller than previously predicted. However,
if $q>q_{\rm crit}$, then the material forced to corotate loses
angular momentum and the orbital period increases. We plot this in the
solid line in Fig.~\ref{qc}.  We note that most observed systems have
a mass ratio $q\lesssim 1$ \citep{ritter03} because there is a
critical mass ratio, that depends on $M_2$, above which the mass
transfer becomes unstable \citep[e.g.][]{smith06}. Hence in most
systems the orbital period will decrease when the effects of a
magnetic field are considered.

We also plot the dashed line for the mass ratio below which the period
change is negative. We see that for the larger magnetic fields it is
possible that the orbital period may actually decrease for all mass
ratios. This effect could be significant even for larger mass
ratios. Frictional angular momentum losses only dominate for
$q\lesssim 0.01$ \citep{shara86} so this new mechanism potentially has
a greater effect on the orbital period change.

\begin{figure*}
\begin{center}
\includegraphics[width=8.4cm]{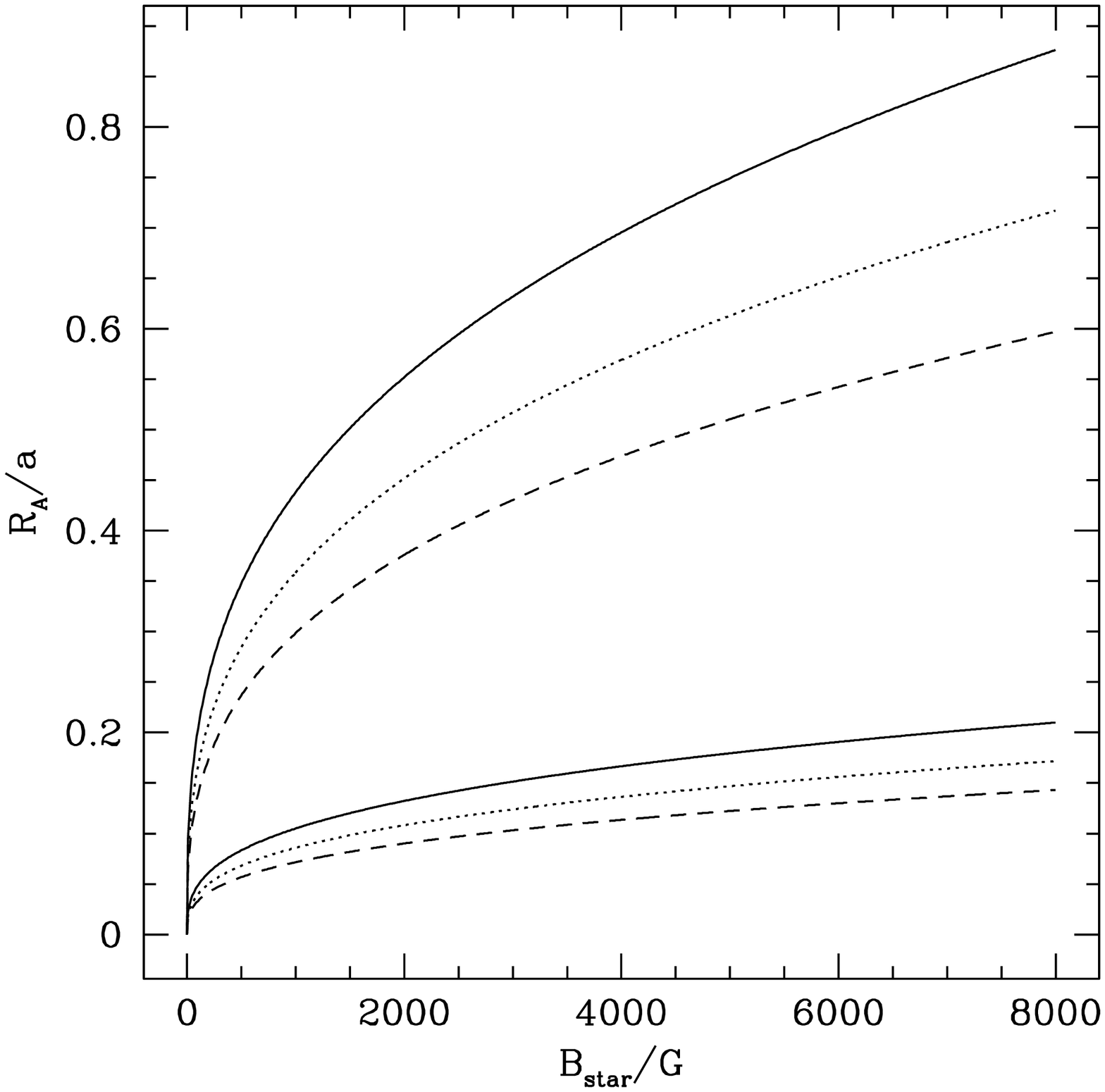}
\includegraphics[width=8.4cm]{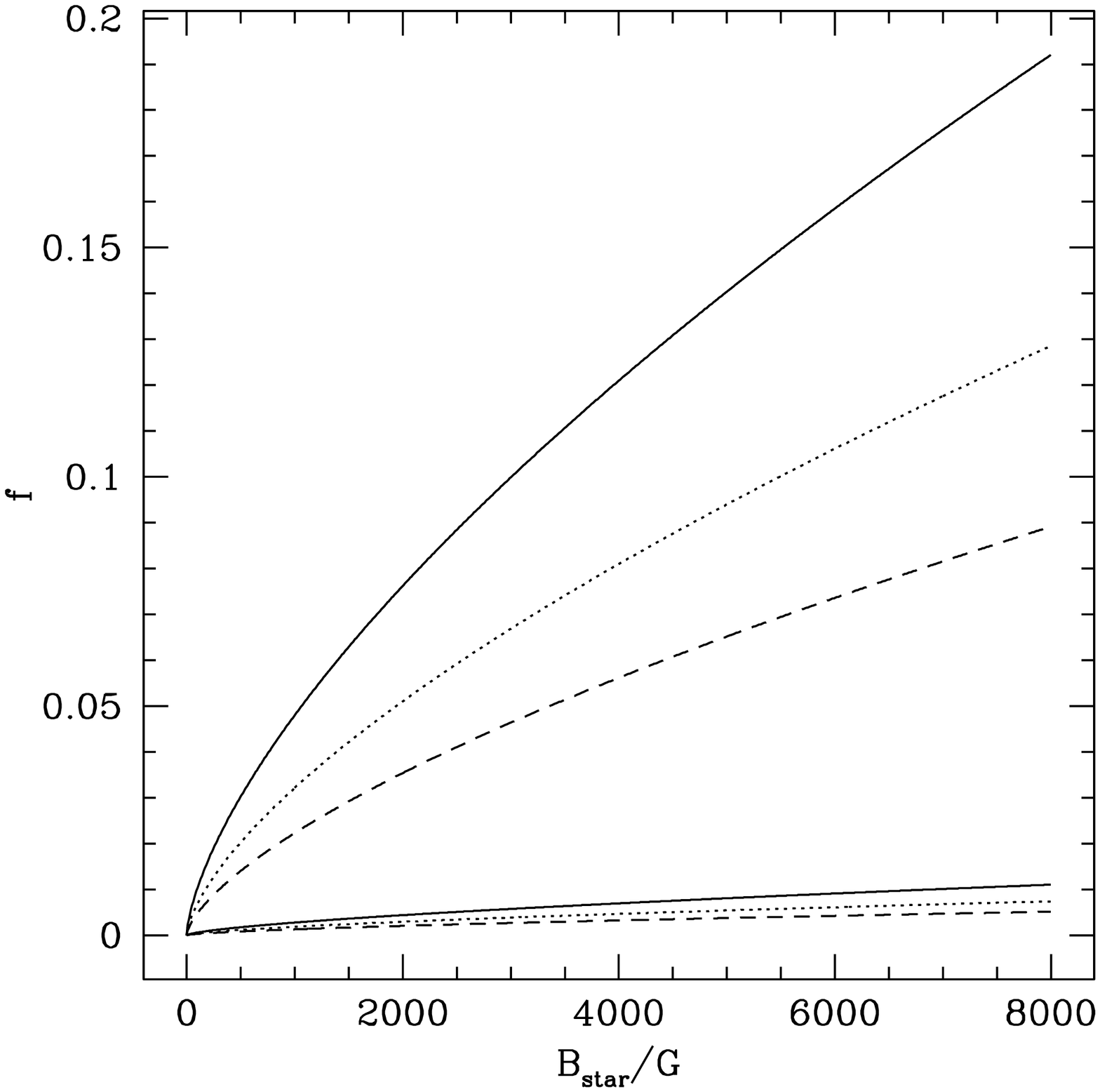}
\end{center}
\caption[]{For a system with the parameters of U Sco. Left: The
  Alfv\'{e}n radius, $R_{\rm A}/a$, as a function of the magnetic
  field strength at the surface of the secondary, $B_{\rm
    star}$. Right: The mass fraction, $f$, that is ejected within the
  Alfv\'{e}n radius of the secondary. The upper three lines are for
  the upper limit on the nova timescale, $\tau$, and the lower three
  lines for the lower limit. In both plots, the solid lines are for
  ejection velocity $u=300\,\rm km\,s^{-1}$, the dotted lines for
  $u=1000\,\rm km\,s^{-1}$ and the dashed lines for $u=3000\,\rm
  km\,s^{-1}$. }
\label{alf}
\end{figure*}

\section{Recurrent Novae}
\label{usco}

There are now ten recurrent novae known in our galaxy \citep[the tenth
  one was discovered last year;][]{pagnotta09} and one system in the
LMC. In this group, U Sco has the fastest decline rate of the light
curve in past outbursts, and the shortest recurrence period
\citep[$11\,\rm yr$ since the last outburst,][]{schaefer01}. It has
outbursts recorded in 1863, 1906, 1917, 1936, 1945, 1969, 1979, 1987,
1999 \citep{schmeer99,schaefer10a} and 2010 \citep{schaefer10b} and others have
likely been missed because of its proximity to the Sun
\citep{schaefer04}.

U Sco has a white dwarf with a mass $M_1=1.55\pm 0.24\,\rm M_\odot$
\citep{thoroughgood01}.  We take the mass to be close to the upper
limit for that of a white dwarf that is accreting matter before a
supernova occurs, so $M_1=1.37\,\rm M_\odot$ \citep{hachisu00a}. This
mass is consistent with the fact that U Sco has frequent
outbursts. The orbital period is $P=1.2305521\,\rm d$
\citep{schaefer90,schaefer95}.  The radius of a non-rotating white
dwarf is given approximately by
\begin{equation}
R_1=7.99 \times 10^8 \left[ \left(\frac{M_{1}}{M_{\rm ch}}\right)^{-\frac{2}{3}}-\left(\frac{M_{1}}{M_{\rm ch}}\right)^\frac{2}{3}\right]^\frac{1}{2} \,\rm cm,
\label{rad}
\end{equation}
where $M_{\rm ch}=1.44\,\rm M_\odot$ is the Chandrasekhar mass
\citep{nauenberg72}, so the radius of the white dwarf in U Sco is
$R_1=0.003\,\rm R_\odot$. 

With equation~(\ref{dm}) we find the mass accumulated before the
outburst to be $\Delta m_1=2.36\times 10^{-6}\,\rm M_\odot$,
consistent with estimates for the 1999 outburst \citep{hachisu00b} and
the 2010 outburst \citep{banerjee10,diaz10}. We assume that all of
this mass is ejected in the outburst. With this we find the ratio
$\Delta m_1/M=1.05\times 10^{-6}$.  The companion to the white dwarf
in the system is a subgiant \citep{schaefer90} with a mass of
$M_{2}=0.88 \,\rm M_\odot$ and a radius of $R_{2}=2.1\,\rm R_\odot$
\citep{thoroughgood01}.  We can take an upper limit of $\tau\approx
3\,\rm months$ \citep[the timescale on which the optical light curve
  drops back to quiescence,][]{matsumoto03} and a lower limit of
$\tau\approx a/u$ (the binary crossing time). We consider a range of
ejection velocities of the material between $u=300\,\rm km\,s^{-1}$
and $3000\,\rm km\,s^{-1}$. With this range of velocities, the lower
limit to the ejection time is in the range $0.5-4\,\rm hr$.

In the left hand side of Fig.~\ref{alf} we plot the Alfv\'{e}n radius
(given in equation~\ref{ra}) for this system for different ejection
velocities. For strong surface magnetic fields of a few thousand
Gauss, it is a significant fraction of the binary separation. On the
right hand side we show the fraction of mass that is ejected within
the Alfv\'{e}n radius (given in equation~\ref{eqf}) of the secondary
star for different mass ejection velocities. Even for the strongest
surface magnetic fields, the fraction of the ejected mass is always a
small fraction of the total mass ejected.

The mass ratio in U Sco is around $q=0.64$. With Fig.~\ref{dadm2} we
see that with a large surface magnetic field, and hence Alfv\'{e}n
radius of the secondary star, the period change during an outburst in
U Sco is significantly reduced. With a strong enough magnetic field
the period change could even be negative (see
Fig.~\ref{qc}). Measurements of the orbital period after the 2010
outburst are strongly encouraged in order to test this mechanism.

\section{Discussion and Conclusions}

We have considered the effect of a strong magnetic field on the
surface of the secondary star on the orbital period change during a
nova outburst. For most systems, the ejected material gains angular
momentum as it couples to the magnetic field. The binary system loses
angular momentum and the period change that results purely from mass
loss decreases.

Our results show that, contrary to expectations, the orbital period in
nova systems could {\it decrease} during outbursts, even in systems in
which it would previously have been expected to increase
\citep[e.g. for $P \gtrsim 8\,\rm hr$,][]{livio91}. Very accurate
determinations of pre- and post-outburst orbital periods in recurrent
nova systems are therefore strongly encouraged. These could provide
meaningful constraints of such quantities as the mass of the accreted
envelope and the secondary's magnetic field.

Magnetic fields of the order of a few kilogauss on the secondary star
have been suggested previously \citep[e.g.][]{meyer96,warner96}. While
such strong fields are typical of magnetic Ap stars, they may be less
common in the secondaries of cataclysmic variables. However, high
magnetic fields have been discussed for cataclysmic variables of
shorter periods by \cite{meintjes06}.  The subgiant companion in U Sco
is expected to be synchronously rotating with the orbit, with a period
of $30\,\rm hr$, which is very fast for a subgiant.  However, studies
of late-type stars show that high fields can be expected for fast
rotators \citep[e.g. ][]{noyes84}.

\section*{Acknowledgements}

We thank Jim Pringle for helpful discussions.


\label{lastpage}
\end{document}